\documentstyle[12pt,fleqn]{article}
\def\soc{{\rm C}_{60}}
\def\rug{{\rm C}_{70}}
\def\c76{{\rm C}_{76}}
\def\c78{{\rm C}_{78}}
\def\c84{{\rm C}_{84}}

\def\la{\langle}
\def\ra{\rangle}
\def\beeq{\begin{equation}}
\def\eneq{\end{equation}}
\def\beeqa{\begin{eqnarray}}
\def\eneqa{\end{eqnarray}}

\addtocounter{section}{-1}
\begin{document}
\begin{center}

{\large {\bf{Optical absorption spectra and geometric effects\\
in higher fullerenes
} } }

\vspace{1cm}

(Running head: {\sl Optical absorption in higher fullerenes})

\vspace{1cm}

{\rm Kikuo Harigaya\footnote[1]{E-mail address: 
harigaya@etl.go.jp.}
and Shuji Abe}\\

\vspace{1cm}

{\sl Physical Science Division,\\
Electrotechnical Laboratory,\\ 
Umezono 1-1-4, Tsukuba, Ibaraki 305, Japan}

\vspace{1cm}

(Received~~~~~~~~~~~~~~~~~~~~~~~~~~~~~~~~~~~)
\end{center}

\vspace{1cm}

\noindent
{\bf Abstract}\\
The optical excitations in $\rug$ and higher fullerenes, 
including isomers of C$_{76}$, C$_{78}$, and C$_{84}$, 
are theoretically investigated.  We use a tight binding 
model with long-range Coulomb interactions, treated by 
the Hartree-Fock and configuration-interaction methods.  
We find that the optical excitations in the energy region 
smaller than about 4 eV have most of their amplitudes at 
the pentagons.  The oscillator strengths of projected 
absorption almost accord with those of the total absorption.  
When the projection is performed on each pentagon and 
pentagon dimers, the resultant spectrum in the low energy 
region is quite different from that of the total absorption.   
The spectral shapes of the total absorption are turned out 
to be determined mainly by the geometrical distributions 
of the pentagons in the fullerene structures.

\mbox{}


\pagebreak

\section{Introduction}

Recently, the fullerene family C$_N$ with hollow cage structures 
has been intensively investigated.  A lot of optical experiments 
have been performed, and excitation properties due to $\pi$-electrons 
delocalized on molecular surfaces have been measured.  For example, 
the optical absorption spectra of $\soc$ and $\rug$ [1,2] have been 
reported, and the large optical nonlinearity of $\soc$ [3,4] has 
been found.  The absorption spectra of higher fullerenes (C$_{76}$, 
C$_{78}$, C$_{84}$, etc.) have also been obtained [5,6].  For 
theoretical studies, we have applied a tight binding model [7] to 
$\soc$, and have analyzed the nonlinear optical properties.  Coulomb 
interaction effects on the absorption spectra and the optical 
nonlinearity have been also studied [8].  We have found that the 
linear absorption spectra of $\soc$ and $\rug$ are well explained 
by the Frenkel exciton picture [9] except for the charge transfer 
exciton feature around the excitation energy 2.8 eV of the $\soc$ 
solids [2].  Coulomb interaction effects reduce the magnitude of 
the optical nonlinearity from that of the free electron calculation 
[8], and thus the intermolecular interaction effects have turned 
out to be important.

In the previous paper [10], we have extended the calculation of 
$\soc$ [9] to one of the higher fullerenes C$_{76}$.  We have 
discussed variations of the optical spectral shape in relation 
to the symmetry reduction from $\soc$ and $\rug$ to C$_{76}$: 
the optical gap decreases and the spectra exhibit a larger number 
of small structures in the dependences on the excitation energy.  
These properties seem to be natural when we take into account of 
the complex surface patterns composed of pentagons and hexagons.  
In order to understand the patterns clearly, the idea of the 
``phason line'' has been introduced [11] using the projection 
method on the honeycomb lattice plane [12].  There are twelve 
pentagons in C$_{76}$.  Six of them cluster on the honeycomb 
lattice, with one hexagon between the neighboring two pentagons.  
There are two groups of the clustered pentagons.  The ``phason line" 
runs as if it divides the two groups.

The purpose of this paper is to investigate relations between 
the optical properties and geometric structures in higher
fullerenes.  The phason lines can characterize the geometries
of fullerenes.  We define the ``pentagonal" carbons as the
atoms located at the points of pentagons.  Then, the carbon
atoms along the phason lines can be regarded as the remaining
atoms after the petangonal carbons are taken out.  We will
calculate the optical absorption spectra at a certain combination
of the pentagonal carbons.  We project wavefunctions to selected
pentagons, and calculate dipole moments using the projected
wavefunctions.  Thus, the contributions from a part of the 
fullerene to the optical spectra can be extracted.

The main results are as follows:\\
(1)  The optical excitations in the energy region smaller 
than about 4 eV have most of their amplitudes at the pentagonal 
carbons.  The oscillator strengths of absorption projected 
onto these carbons almost accord with those of the total 
absorption.\\
(2)  When the projection is performed on a smaller set of 
the pentagonal carbons, the resultant spectrum in the low 
energy region is quite different from that of the total 
absorption.  The structures in the absorption cannot be 
decomposed into contributions from subsets of pentagonal 
carbons.\\
(3) The contributions from pentagon dimers in several kinds 
of higher fullerenes (e.g., two isomers of C$_{78}$ and C$_{84}$)
are compared.  We find that small peak structures are mutually 
different.  The spectral shapes of the total absorption are 
mainly determined by the geometrical distributions of the 
pentagons in the fullerene structures.

In the next section, our tight-binding model is introduced 
and the calculation method is explained.  Sections 3 and 4 
are devoted to numerical results and discussion about the 
properties of the absorption spectra.  The paper is closed 
with a summary in section 5.

\section{Model}

We use the following Hamiltonian:
\beeq
H = H_0 + H_{\rm int}.
\eneq
The first term of eq. (1) is the tight binding model:
\beeq
H_0 = - t \sum_{\langle i,j \rangle, \sigma} 
(c_{i,\sigma}^\dagger c_{j,\sigma} + {\rm h.c.}),
\eneq
where $t$ is the hopping integral and $c_{i,\sigma}$ is an 
annihilation operator of a $\pi$-electron with spin $\sigma$
at the $i$th carbon atom of the fullerene.  It is assumed 
that $t$ does not depend on the bond length, because main 
contributions come from excitonic effects due to the strong 
Coulomb potential.  The results do not change so largely 
if we consider changes of hopping integrals by bond distortions.  
We assume the following form of Coulomb interactions among 
$\pi$-electrons:
\beeqa
H_{\rm int} &=& U \sum_i 
(c_{i,\uparrow}^\dagger c_{i,\uparrow} - \frac{1}{2})
(c_{i,\downarrow}^\dagger c_{i,\downarrow} - \frac{1}{2}) \nonumber \\
&+& \sum_{i \neq j} W(r_{i,j}) 
(\sum_\sigma c_{i,\sigma}^\dagger c_{i,\sigma} - 1)
(\sum_\tau c_{j,\tau}^\dagger c_{j,\tau} - 1),
\eneqa
where $r_{i,j}$ is the distance between the $i$th and $j$th atoms and
\beeq
W(r) = \frac{1}{\sqrt{(1/U)^2 + (r/r_0 V)^2}}
\eneq
is the Ohno potential.  The quantity $U$ is the strength of 
the on-site interaction, $V$ means the strength of the long-range
Coulomb interaction, and $r_0$ is the average bond length.

The model is treated by the Hartree-Fock approximation and the 
single excitation configuration interaction method, as was used 
in the previous papers [9,10].   In ref. 9, we have varied the 
parameters of the Coulomb interactions, and have searched for 
the data which reproduce overall features of experiments of 
$\soc$ and $\rug$ in solutions.  We have found that the common 
parameters, $U = 4t$ and $V = 2t$, are reasonable.  Thus, we use 
the same parameter set for higher fullerenes.  The quantity $t$ 
is about 2 eV as shown in ref. 9.  The Coulomb interaction strengths 
depend on the carbon positions.  We use the lattice coordinates 
which are obtained by the public program ``FULLER" [13,14].  The 
optical spectra become anisotropic with respect to the orientation 
of the molecule against the electric field of light, as reported 
in the free electron model (H\"{u}ckel theory) [15].  We obtain 
numerical optical absorption spectra by summing the data of three 
cases, where the electric field of light is along the $x$-, $y$-, 
and $z$-axes.

We use a projection operator in order to extract contributions
to the optical spectra from a certain part of fullerenes.  
If we write the a projection operator to a part of lattice site
set as $P$, the oscillator strength between the ground state 
$|g\rangle$ and the excited state $|\kappa \rangle$ is written:
\beeqa
f_{\kappa,x} &=& E_\kappa [ | \la \kappa | P x P | g \ra |^2 \nonumber \\
&+& | \la \kappa | (1-P) x (1-P) | g \ra |^2 \nonumber \\
&+& \la g | P x P | \kappa \ra \la \kappa | (1-P) x (1-P) | g \ra \nonumber \\
&+& \la g | (1-P) x (1-P) | \kappa \ra \la \kappa | P x P | g \ra ],
\eneqa
where $E_\kappa$ is the excitation energy, and the electric
field is parallel with the $x$-axis.  In eq. (5), the first term
is the contribution from the projected part, and the three other
terms are the remaining part.  The total optical absorption
is calculated by the formula:
\beeq
\sum_\kappa \rho(\omega - E_\kappa) 
(f_{\kappa,x} + f_{\kappa,y} + f_{\kappa,z}),
\eneq
where $\rho(\omega) = \gamma/[\pi(\omega^2+\gamma^2)]$ 
is the Lorentzian distribution of the width $\gamma$.  
The projected absorption is calculated by eqs. (5) and (6).  
The projected part does not satisfy a sum rule.  So, this 
results in a singularity where excitation energy is large.  
We will discuss the optical spectra in the energy region
far from the singularity.

\section{Optical absorption in C$_{\bf 70}$ and C$_{\bf 76}$}

Figure 1 shows the molecular structures and optical spectra
of the $\rug$ molecule and C$_{76}$ with the $D_2$ symmetry, 
which have been found in experiments.  The black atoms are 
the carbons along the phason lines.  The hatched circles are 
the pentagonal carbons.  In C$_{70}$, the phason line runs 
along the ten carbons which are arrayed like a belt around 
the molecule.  In C$_{76}$, the phason line is located almost 
along the outer edge of the molecule of Fig. 1(b).  The total 
optical absorption is shown by the bold line, and the 
absorption from all the pentagonal carbons is shown by the 
thin line.  We find that the optical excitations in the energy 
region lower than 2$t$ are almost composed of the excitations 
at the pentagonal sites.  This property is common to $\rug$
and $D_2$-C$_{76}$, and also to the $T_d$-C$_{76}$ for which
the calculated data are not shown.  In higher energy regions,
the thin lines give relatively larger oscillator strengths,
but this is an artifact of the projected wavefunctions.  The 
absorption spectra calculated from the projected wavefunctions
do not satisfy the sum rule, i.e., the area between the abscissa
and the curve does not become constant regardless of the 
excitation wavefunctions.  The similar artifact will be found
in the figures shown afterwards.  We believe that the projected
optical absorption spectra are reliable in low energy regions
only.  Therefore, we limit our comparison of the spectra to
the energy region lower than about $2t \sim 4$ eV.

In $\soc$, the edges of the pentagons are the long bonds,
and the bonds between the neighboring hexagons are short
bonds.  The wavefunctions of the fivefold degenerate 
highest-occupied-molecular-orbital (HOMO) have the bonding 
property, and that the threefold degenerate 
lowest-unoccupied-molecular-orbital (LUMO) has the 
antibonding property.  As the carbon number 
increases, hexagons are inserted among pentagons.  The 
wavefunctions near the LUMO of the higher fullerenes still 
have the antibonding properties, thus they tend to have 
large amplitudes along the edges of pentagons which have 
the characters like long bonds of $\soc$.  Recently, the 
bunching of the six energy levels higher than the LUMO  
has been discussed in the extracted higher fullerenes [16].  
The wavefunctions near the LUMO distribute on the pentagons.  
This fact can be understood as the properties characteristic 
to antibonding orbitals.  As the excited electron mainly 
distributes at the pentagonal carbons, the electron-hole 
excitation has large amplitudes at these pentagons.  Thus,
the oscillator strengths of the low energy excitations are
mainly determined by wavefunctions at the pentagonal carbons.
This is the reason why the projected absorptions nearly
accord with the total absorptions in the energy regions
smaller than about 2$t$.

If the projections are performed onto each pentagon, we can
know contributions to optical spectra from the projected carbon
sites.  We would like to look at this feature, for example,
in $D_2$-C$_{76}$.  There are three carbon atoms, which are
not equivalent with respect to symmetries, in this isomer.
These pentagons are indicated by the symbols, A-C, in Fig. 2(a).
The projected absorption spectra are shown by thin curves,
superposed with the total absorption in Figs. 2 (b-d).
The projected absorption is multiplied by the factor 12, 
in order to compare with the total absorption.  We find 
that the projected absorption exhibits small structures 
in the energy region smaller than $2t$.  The structures 
depend on the kind of carbons.  The spectral shapes and 
oscillator strengths are much far from those of the total 
absorption.  It would be difficult to assign experimental 
features of the total absorption with a set of the limited 
number of carbon atoms.  The excitation wavefunctions at 
the twelve pentagons give rise to the shape of the 
absorption spectra totally.

\section{Optical absorption in C$_{\bf 78}$ and C$_{\bf 84}$}

Figures 3 and 4 show the molecular structures and the 
calculated optical absorption spectra of the extracted 
isomers of C$_{78}$ and C$_{84}$, respectively.  The 
notations in Figs. 3 and 4 are the same as in Fig. 1.  
The property, that the low energy excitations have most 
of their amplitudes at the twelve pentagons, is seen in 
these five isomers.  The bold line and the thin line 
almost agree in the energy region lower than about 
$2t \sim 4$ eV.  It is of course that the agreement 
becomes a little bit worse from that in $\rug$ and 
C$_{76}$ of Fig. 1, because the number of atoms along
the phason lines increases as the fullerenes become higher.
But, the similar wavefunction properties seem to persist
in the calculated fullerenes, C$_{78}$ and C$_{84}$.

The pentagon dimers exist in the regions surrounded by 
the phason lines in the two $C_{2v}$ isomers of C$_{78}$
[Figs. 3 (a) and (b)] and also in the $D_{2d}$ isomer
of C$_{84}$ [Fig. 4 (a)].  These three molecules have
pentagon dimer structures commonly.  The dimers are shown
by the black pentagons in molecular structures of Fig. 5.  
It is of some interests to look at whether these pentagon 
give rise to similar contributions to the optical spectra
or not.  The data of the thin lines are multiplied by the 
factor 6, in order to compare with the total absorption.  
We find that small peak structures are mutually different:
for a typical example, the large feature around the energy 
1.5$t$ represented by the thin line in Fig. 5(c) is not present 
in Figs. 5(a) and (b).  We thus conclude that the spectral 
shapes of the total absorption are mainly determined by the 
geometrical distributions of the pentagons in the fullerene 
structures.

\section{Summary}

We have studied optical excitations in $\rug$ and higher 
fullerenes, including isomers of C$_{76}$, C$_{78}$, and 
C$_{84}$.  We have used a tight binding model with 
long-range Coulomb interactions, and have treated by the 
Hartree-Fock and CI methods.  We have found that the optical 
excitations in the energy region smaller than about 4 eV 
have most of their amplitudes at the pentagonal carbons.  
Thus, the oscillator strengths of absorption projected 
onto these carbons almost accord with those of the total 
absorption.  When the projection is performed on a smaller 
set of the pentagonal carbons, for example, on each pentagon, 
the resultant spectrum in the low energy region is quite 
different from that of the total absorption.  Therefore, 
the structures in the absorption cannot be decomposed into 
contributions from subsets of pentagonal carbons.

\mbox{}

\noindent
{\bf Acknowledgements}\\
The authors acknowledge useful discussion with 
Y. Achiba, S. Saito, M. Fujita, and M. Yoshida.

\pagebreak
\begin{flushleft}
{\bf References}
\end{flushleft}

\noindent
$[1]$ J. P. Hare, H. W. Kroto and R. Taylor, Chem. Phys. Lett. 
177 (1991) 394.\\
$[2]$ S. L. Ren, Y. Wang, A. M. Rao, E. McRae, J. M. Holden, 
T. Hager, K. A. Wang, W. T. Lee, H. F. Ni, J. Selegue 
and P. C. Eklund, Appl. Phys. Lett. 59 (1991) 2678.\\
$[3]$ J. S. Meth, H. Vanherzeele and Y. Wang, Chem. Phys. Lett.
197 (1992) 26.\\
$[4]$ Z. H. Kafafi, J. R. Lindle, R. G. S. Pong, F. J. Bartoli,
L. J. Lingg and J. Milliken, Chem. Phys. Lett. 188 (1992) 492.\\
$[5]$ R. Ettl, I. Chao, F. Diederich and R. L. Whetten,
Nature 353 (1991) 149.\\
$[6]$ K. Kikuchi, N. Nakahara, T. Wakabayashi, M. Honda, H. Matsumiya,
T. Moriwaki, S. Suzuki, H. Shiromaru, K. Saito, K. Yamauchi,
I. Ikemoto and Y. Achiba, Chem. Phys. Lett. 188 (1992) 177.\\
$[7]$ K. Harigaya and S. Abe, Jpn. J. Appl. Phys. 31 (1992) L887.\\
$[8]$ K. Harigaya and S. Abe, J. Lumin. 60\&61 (1994) 380.\\
$[9]$ K. Harigaya and S. Abe, Phys. Rev. B 49 (1994) 16746.\\
$[10]$ K. Harigaya, Jpn. J. Appl. Phys. 33 (1994) L786.\\
$[11]$ M. Fujita, Fullerene Science and Technology, 1 (1993) 365.\\
$[12]$ M. Fujita, R. Saito, G. Dresselhaus and M. S. Dresselhaus,
Phys. Rev. B 45 (1992) 13834.\\
$[13]$ M. Yoshida and E. \={O}sawa, Proc. 3rd IUMRS
Int. Conf. Advanced Materials, 1993.\\
$[14]$ M. Yoshida and E. \={O}sawa, 
The Japan Chemistry Program Exchange, Program No. 74.\\
$[15]$ J. Shumway and S. Satpathy, Chem. Phys. Lett. 211 (1993) 595.\\
$[16]$ S. Saito, S. Okada, S. Sawada and N. Hamada,
Phys. Rev. Lett. 75 (1995) 685.\\

\pagebreak

\begin{flushleft}
{\bf Figure Captions}
\end{flushleft}

\mbox{}

\noindent
Fig. 1.  Molecular structures and theoretical optical
spectra of (a) $D_{5d}$-$\rug$ and (b) $D_2$-C$_{76}$.
In the molecules, the black atoms are along the phason
lines, and the hatched atoms are the pentagonal carbons.
In the absorption spectra, the bold line is the total
absorption, and the thin line is the absorption by 
the wavefunctions projected on the twelve pentagons.
The units of the abscissa are taken as arbitrary, and
the energy is scaled by $t$.  The parameters are
$U=4t$, $V=2t$, and $\gamma = 0.06t$.

\mbox{}

\noindent
Fig. 2.  (a) The molecular structure of $D_2$-C$_{76}$.  The 
symbols, A-C, indicate the symmetry nonequivalent pentagons.
The figures, (a), (b), and (c), compare the absorption 
projected on one of the three pentagons with the total 
absorption.  The bold line is the total absorption, and 
the thin line is the projected absorption.  The units of 
the abscissa are taken as arbitrary, and the energy is 
scaled by $t$.  The data of the thin line are multiplied
by the factor 12.  The parameters are $U=4t$, $V=2t$, 
and $\gamma = 0.06t$.

\mbox{}

\noindent
Fig. 3.  Molecular structures and theoretical optical
spectra of (a) $C_{2v}$-C$_{78}$, (b) the other 
$C_{2v}$-C$_{78}$, and (c) $D_3$-C$_{78}$.
In the molecules, the black atoms are along the phason
lines, and the hatched atoms are the pentagonal carbons.
In the absorption spectra, the bold line is the total
absorption, and the thin line is the absorption by 
the wavefunctions projected on the twelve pentagons.
The units of the abscissa are taken as arbitrary, and
the energy is scaled by $t$.  The parameters are
$U=4t$, $V=2t$, and $\gamma = 0.06t$.

\mbox{}

\noindent
Fig. 4.  Molecular structures and theoretical optical
spectra of (a) $D_{2d}$-C$_{84}$ and (b) $D_2$-C$_{84}$.
In the molecules, the black atoms are along the phason
lines, and the hatched atoms are the pentagonal carbons.
In the absorption spectra, the bold line is the total
absorption, and the thin line is the absorption by 
the wavefunctions projected on the twelve pentagons.
The units of the abscissa are taken as arbitrary, and
the energy is scaled by $t$.  The parameters are
$U=4t$, $V=2t$, and $\gamma = 0.06t$.

\mbox{}

\noindent
Fig. 5.  Molecular structures and theoretical optical
spectra of (a) $C_{2v}$-C$_{78}$, (b) the other 
$C_{2v}$-C$_{78}$, and (c) $D_{2d}$-C$_{84}$.
In the molecules, the pentagon dimers are shown by 
the black pentagons.  The each figure compares the 
absorption projected on the pentagon dimer with the total 
absorption.  The bold line is the total absorption, and 
the thin line is the projected absorption.  The units of 
the abscissa are taken as arbitrary, and the energy is 
scaled by $t$.  The data of the thin line are multiplied
by the factor 6.  The parameters are $U=4t$, $V=2t$, and 
$\gamma = 0.06t$.

\end{document}